\documentclass[aip,jcp,preprint,groupedaddress,raggedbottom,floatfix]{revtex4-1} 

\usepackage{bm}
\usepackage{amsmath}
\usepackage{amsfonts}
\usepackage{amssymb}
\usepackage{graphicx}
\usepackage{dcolumn}
\usepackage{bm}
\usepackage{setspace} 
\usepackage{color}
\usepackage{ragged2e} 

\begin{document}

\title{Remarkably low-energy one-dimensional fault line defects in single-layered phosphorene}

\author{Woosun \surname{Jang}}
\affiliation{Global E$^{3}$ Institute and Department of Materials Science and Engineering, Yonsei University, Seoul 120-749, Korea}
\author{Kisung \surname{Kang}}
\affiliation{Global E$^{3}$ Institute and Department of Materials Science and Engineering, Yonsei University, Seoul 120-749, Korea}
\author{Aloysius \surname{Soon}}
\email[Corresponding author. E-mail: ]{aloysius.soon@yonsei.ac.kr}
\affiliation{Global E$^{3}$ Institute and Department of Materials Science and Engineering, Yonsei University, Seoul 120-749, Korea}

\date{\today}

\begin{abstract}
Systematic engineering of atomic-scale low-dimensional defects in two-dimensional nanomaterials is a promising way to modulate the electronic properties of these nanomaterials. Defects at interfaces such as grain boundaries and line defects can often be detrimental to technologically important nanodevice operations and thus a fundamental understanding of how such one-dimensional defects may have an influence on its physio-chemical properties is pivotal to optimizing their device performance. Of late, two-dimensional phosphorene has attracted much attention due to its high carrier mobility and good mechanical flexibility. In this study, using density-functional theory, we investigate the temperature-dependent energetics and electronic structure of a single-layered phosphorene with various fault line defects. We have generated different line defect models based on a fault method, rather than the conventional rotation method. This has allowed us to study and identify new low-energy line defects, and we show how these low-energy line defects could well modulate the electronic band gap energies of single-layered two-dimensional phosphorene -- offering a range of metallic to semiconducting properties in these newly proposed low-energy line defects in phosphorene.

\end{abstract}

\maketitle

\clearpage
\newpage

\begin{flushleft}
{\bf Graphical Abstract for Table-of-Content:} 
\end{flushleft}
\justify
One-dimensional (1D) line defects are ubiquitous in polycrystalline two-dimensional nanomaterials. Various structures of phosphorene with 1D fault line defects were designed and studied using first-principles density-functional theory calculations. We identified new low-energy fault line defects (i.e. even much lower fault defect formation energy when compared to those found on other two-dimensional (2D) nanomaterials) and showed how these low-energy fault line defects may help to tune the phosphorene-based nanodevices' performance.
\vspace{2.5mm}
\begin{center}
\includegraphics[width=0.95\textwidth]{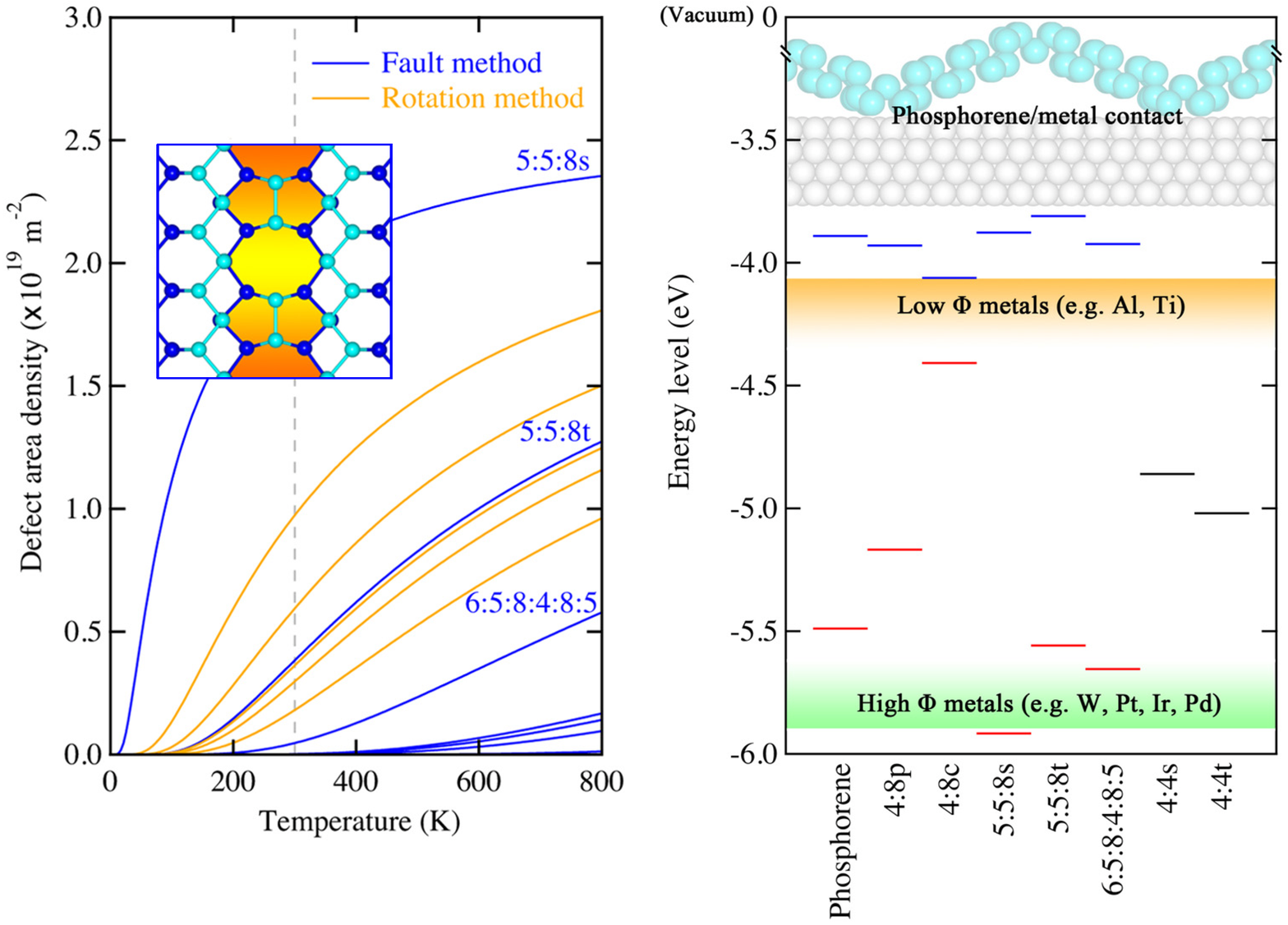}
\end{center}

\clearpage
\newpage

\section{Introduction}
In real materials, a perfect crystal is an ``ennoblement", i.e. an idealized state. It is fundamentally important to understand the imperfections in real crystalline materials, and studying of the science behind these crystal irregularities is critical to functional materials design and engineering.\cite{Zou2014,Zou2015,JWang2015,Carlsson2011} It is the feasibility of creating geometrically-imperfect materials that allows us to ``custom-make" new materials with assorted materials functions and properties that modern technological devices desire. In this sense, one of the most important characteristics of a material's microstructure are these very imperfections that are exploited for the engineering design of new materials.\cite{Zou2015,JWang2015}

Of late, two-dimensional (2D) nanomaterials have attracted a lot of interest from the optoelectronics community.\cite{Zou2014,Zou2015,Zhou2013,Wang2012} Since the successful isolation of graphene (from its bulk form -- graphite)\cite{Novoselov2004,Novoselov2005} and the 2D metal dichalcogenides (e.g. MoS$_2$),\cite{Wang2012,Jariwala2014} the high carrier mobilities of these 2D nanomaterials has sparked great expectations as the most probable candidates to supersede and replace traditional semiconductors in key nanodevices e.g. the field-effect transistor (FET).\cite{Schwierz2015,Li2014,HLiu2014}

However, after much enthusiasm and anticipation, it has become clear that graphene, which is a semi-metal, does not possess a band gap needed for proper device operations while many of these 2D metal dichalcogenides suffer from the existence of deep defect states in the middle of the band gap.\cite{Wang2012,Jariwala2014} These defects are often a result of line defects at the grain boundaries (GBs) which can act as unwanted electron-hole recombination centers that hinders and degrades nanodevice performance.\cite{YLiu2014}

\begin{figure}[b!]
\centering
\includegraphics[width=0.65\textwidth]{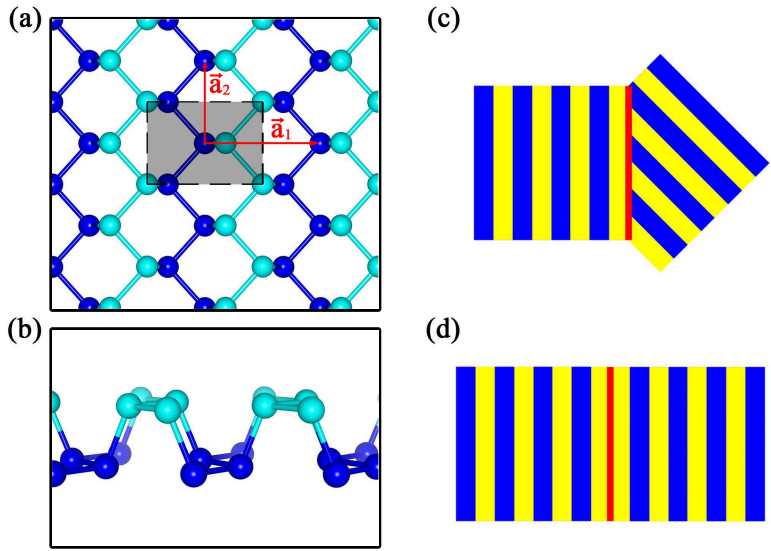} 
\caption{(Color online) (a) Top- and (b) side-views of the atomic geometry of a single-layered phosphorene. To aid viewing, atoms at different vertical heights are discriminated by different shades of color. The grayed area in (a) denotes the $p(1\times1)$ Wigner-Seitz cell. To generate line/fault defects, namely (c) the {\it rotation method} (upper panel) and (d) the {\it fault method} (lower panel) can be employed.}
\label{figure1}
\end{figure}

Yet another recently discovered 2D element -- phosphorene,\cite{Brent2014,Buscema2014,Koenig2014} the 2D analogue of bulk phosphorous, has been placed in the spotlight for its high carrier mobility and a tunable direct electronic band gap.\cite{Buscema2014,Koenig2014,Cai2014,HLiu2014,Kou2015} 2D phosphorene, which has puckered repeating unit geometry (as illustrated in Figs.\,\ref{figure1}a (top-view) and \ref{figure1}b (side-view)), is also found to have superior mechanical flexibility, as compared to other 2D nanomaterials.\cite{Wei2014} 

To date, much attention has been given to point and substitutional defects in 2D nanomaterials. Here, we focus on one-dimensional (1D) line defects where these prevalent (but less studied) defects may be found at the interfaces of such two-dimensional nanomaterials (e.g. grain boundaries and micro-faceted nanocrystals). 

A detailed discussion of electronically benign point and line defects in phosphorene was recently reported by Liu {\it et al.},\cite{YLiu2014} including a survey of the role of GB orientation on the overall electronic structure of the defected system. Here they found that, in contrary to the 2D metal dichalcogenides, these one-dimensional (1D) defects at the GBs do not induce mid-gap defect trap states, and further show that the carrier type and concentration of this recently identified 2D phosphorene could be controlled via selective chemical doping.

Conventionally in literature, two different methods have been proposed to generate line defects in these 2D nanosystems.\cite{YLiu2014,Singh2014} First of all, one may apply the so-called {\it rotation method},\cite{YLiu2014,Liu2012} whereby one rotates part of the 2D nanostructure while keeping the remaining part fixed, i.e. generating an ``angled" line defect. Upon changing the degrees of rotation, different line defect structures can be formed with varying contact angles between the two rotated parts (as in Fig.\,\ref{figure1}c). 

Alternately, one might adopt the so-called {\it fault method}, following after the recent theoretical study of line defects in both graphene and hexagonal boron nitride ($h$-BN).\cite{Singh2014,Kahaly2008} Here, two different approaches to create line faults in 2D nanomaterials were suggested, namely the stacking fault method and the growth fault method, respectively. In essence, the stacking fault method involves the addition of an extra row of atoms in the middle of the normal stacking sequence, whereas the growth fault method removes one row of atoms in the normal stacking sequence (i.e. creating a ``missing-row". Refer to Fig.\,\ref{figure1}d).

We stress that the defect structures obtained by both {\it rotation method} and {\it fault method} are atomically different, and thus a comparative study of both methods will further our understanding of various one-dimensional line defects in phosphorene. In this work, we used first-principles density-functional theory calculations to discuss the energetics and electronic structure of these 1D fault line defects, and compare with those reported in Ref.\,\onlinecite{YLiu2014} as well as other 2D nanomaterials like graphene and $h$-BN.

\section{Computational setup and methodology}
In this work, periodic density-functional theory (DFT) calculations are performed using the projector-augmented wave (PAW) method, as implemented in the Vienna \textit{Ab initio} Simulation Package (VASP).\cite{Kresse1993,Kresse1994,Kresse1996} The generalized gradient approximation (GGA) to the exchange-correlation functional due to Perdew, Burke and Ernzerhof (PBE) is used.\cite{Kresse1996,Kresse1999,Perdew1996} 
A planewave kinetic energy cutoff of 500\,eV is applied and the irreducible Brillouin zone integrations are performed using a $\Gamma$-centered {\bf k}-point mesh of 4$\times$3$\times$1 for the $p(3\times4)$ supercell (with 96 atoms) and a 2$\times$6$\times$1 {\bf k}-mesh for the $p(7\times2)$ supercell (with 104 atoms). To accurately describe the electronic structures, HSE06 hybrid functional calculations are performed to obtain density-of-states (DOS) and electronic work function with the same {\bf k}-meshes used in PBE calculations. To minimize the spurious interaction between repeating phosphorene layers in the out-of-plane direction (i.e. the $z$-direction), a vacuum region of 15\,{\AA} is applied along the $z$-direction of supercell. Also, in the effort to minimize the lateral interactions between each periodic image of the line defects in these phosphorene layers, a defect-defect distance of at least 9\,{\AA} is ensured. All energies and forces are converged to less than 20\,meV and 0.02\,eV/{\AA}, respectively.

\section{Atomic structures of one-dimensional line defects}
\begin{figure}[tb!]
\centering
\includegraphics[width=0.75\textwidth]{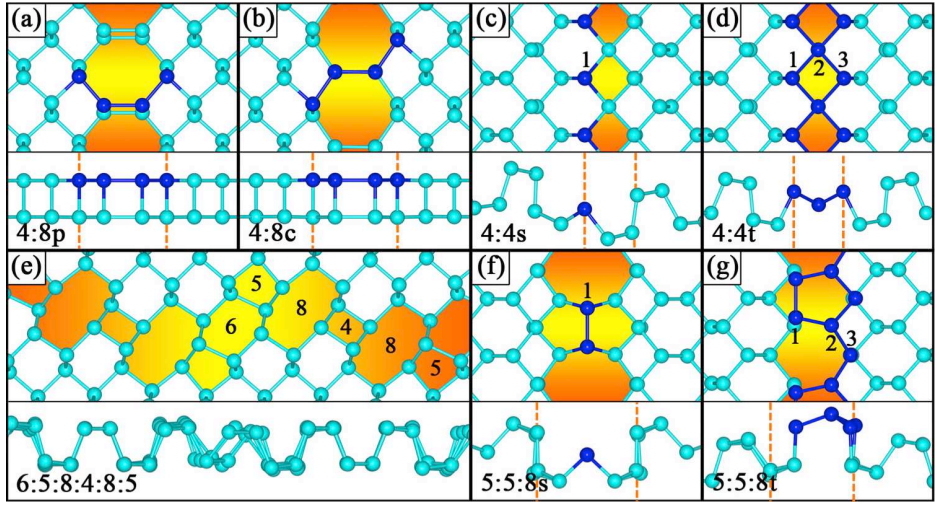} 
\caption{(Color online) Atomic structures of one-dimensional line defects in phosphorene:(a) and (b) show the top-and side-views of the defect geometries of 4:8p and 4:8c, respectively, while (c) and (d) show that of 4:4s and 4:4t, accordingly. The top- and side-views of the atomic structure of 6:5:8:4:8:5 are depicted in (e), whereas that of 5:5:8s and 5:5:8t are shown in (f) and (g), respectively. The line defect region is shaded in orange, while the essential atoms of each line defect is shown in a darker blue to aid viewing.}
\label{figure2}
\end{figure}

\subsection{One-dimensional line defects by the stacking fault method}
Due to the puckered atomic geometry of a single-layered phosphorene, there exists two distinct ways to add an additional row of atoms in pristine phosphorene, along the so-called ``armchair" direction. On one hand, an additional row of atoms can be added with the same stacking sequence as found in pristine phosphorene. This results in the 4:8p line defect structure -- a {\it parallel} (hence, p) puckered structure with sequential rectangular (hence, 4) and octagonal (hence, 8) ring pair along this fault line, as in Fig.\,\ref{figure2}a. 

On the other hand, a single atomic row can be inserted with a reversed ``up-and-down"-like structure, owing to its puckered nature. Here, the 4:8c line defect structure is obtained -- a {\it crossing} (hence, c) puckered structure with a sequential rectangular ring and octagonal ring pair along the fault line, as shown in Fig.\,\ref{figure2}b. 

For both 4:8p and 4:8c, minimal structural relaxation is observed. In the case of 4:8p, the calculated distance between repeated atomic rows is slightly increased from 2.21\,{\AA} to 2.25\,{\AA}, while for the 4:8c structure, a slight outward relaxation of the rectangular ring towards the vacuum is observed, and the distance between repeated atomic rows is also increased further to 2.36\,{\AA}.

\subsection{One-dimensional line defects by the growth fault method}
Similarly to the stacking fault method, due to its puckered nature, the growth fault method presents two different ways of removing a single atomic row from the pristine phosphorene layer, along the so-called ``zigzag" direction. Thus, the two adjacent planes will then have two distinct ways of attachment across the fault. In particular, the two planes could either be directly connected, or by flipping just one plane by $180^{\circ}$ about the fault line before connecting to the other. For the latter case, this means that the uppermost atomic layer now matches that of the bottom-most layer, and vice versa.

From these considerations, two different initial growth fault line defect structures, namely 4:4t and 4:4s are formed, respectively. Specifically, in the case of 4:4s, upon removing one of uppermost atomic row, a single (hence, s) atomic row remains in that layer along the fault line with sequential rhombus rings (hence, 4 as seen in Fig.\,\ref{figure2}c), while for 4:4t, due to the vertical flipping over of one plane, there now exist three (hence, t) atoms in the same layer along the fault line with sequential rhombus rings (hence, 4 as in Fig.\,\ref{figure2}d).

In addition to these 4:4-type line defects, Singh {\it et al.} proposed two other 4:4-derived families of line defects for $h$-BN and graphene, specifically the 6:5:8:4:8:5 and 5:5:8 line defect structures.\cite{Singh2014} In accordance with Ref.\,\onlinecite{Singh2014}, we have also constructed similar line defect structures for single-layered phosphorene -- 6:5:8:4:8:5, 5:5:8s, and 5:5:8t, respectively. Following the naming convention for the line defects mentioned above, as shown in Fig.\,\ref{figure2}e, the 6:5:8:4:8:5 structure has as many $n$th-sided polygons joining sequentially along the line defect as listed accordingly. For 5:5:8s (as in Fig.\,\ref{figure2}f), it shows two slightly folded pentagonal ring paired with one octagonal ring, and a single  outermost top atom exposed along the fault line, which dimerizes with the neighboring atom (i.e. forming P-P pairs) after geometry relaxation. It is worth noting that this structure relaxes to a rather non-planar 2D structure, exhibiting an bent angle of $136.35^{\circ}$. For the 5:5:8t structure in Fig.\,\ref{figure2}g, it has three outermost atoms along the fault line defect, with one almost flat pentagonal ring and one almost $90^{\circ}$-folded pentagonal ring paired with a similarly folded octagonal ring.

Essentially, many of these stacking fault line defects maintain their perfect planar geometry (See Figs.\,\ref{figure2}a and \ref{figure2}b) while those due to growth faults relax to a non-planar geometry with much surface rumpling and wrinkling (as seen in Figs.\,\ref{figure2}c to \ref{figure2}g). This off-planar buckling behavior has been reported for similar multi-center defects in other 2D nanomaterials (e.g. large-angle grain-boundary defects in graphene and $h$-BN).\cite{Singh2014,Liu2012,Yazyev2010} 

\section{Energetics of one-dimensional line defects}
\begin{table}[tb!]
  \centering
\caption{Line defect formation energy ($E^{\rm{LD}}$), band gap ($E_{\rm{g}}$), and work function ($\Phi$) of 1D line defects (LDs) in various 2D nanomaterials: Phosphorene, $h$-BN, and graphene. }
  	\begin{ruledtabular}
    \begin{tabular}{cccccc}
    \noalign{\vskip 1mm}
	Type of LDs			& System		& $E^{\rm{LD}}$ (eV/{\AA}) & $E_{\rm{g}}$	(eV) & $\Phi$ (eV) \\
	\noalign{\vskip 1mm}
	\hline
	\noalign{\vskip 1mm}
	Pristine			& Phosphorene\footnotemark[1]	& 0.00 &	 1.60 & 5.44 					\\ 
	\noalign{\vskip 1mm}
	\hline
	\noalign{\vskip 1mm}
	4:8p				& Phosphorene\footnotemark[1]	& 0.39 & 1.24 & 5.06						\\ 
	4:8c				& Phosphorene\footnotemark[1]	& 0.72 & 0.35 & 4.33						\\
	4:8					& $h$-BN\footnotemark[2] 			& 0.48 & -	 & -	\\
	4:8					& Graphene\footnotemark[2] 			& 0.74  & - & -		\\
	\noalign{\vskip 1mm}
	\hline
	\noalign{\vskip 1mm}
	4:4s				& Phosphorene\footnotemark[1]	& 0.37 & Metallic & 4.86						\\
	4:4t				& Phosphorene\footnotemark[1]	& 0.45 & Metallic & 5.02						\\
	4:4					& $h$-BN\footnotemark[2] 			& 1.87 & -	& -	\\
    \noalign{\vskip 1mm}
	\hline
	\noalign{\vskip 1mm}
	5:5:8s				& Phosphorene\footnotemark[1]	& 0.01 & 2.03 & 5.86						\\
	5:5:8t				& Phosphorene\footnotemark[1]	& 0.10 & 1.75 & 5.46						\\
	5:5:8 N-N			& $h$-BN\footnotemark[2] 			& 0.77 & -	& -	\\
	5:5:8 B-B			& $h$-BN\footnotemark[2] 			& 0.65 & -	& -	\\
	5:5:8				& Graphene\footnotemark[2] 		& 0.52 & -	& -	\\
	\noalign{\vskip 1mm}
	\hline
	\noalign{\vskip 1mm}
	6:5:8:4:8:5			& Phosphorene\footnotemark[1]	& 0.20 & 1.74 & 5.59						\\
	6:5:8:4:8:5 N-N		& $h$-BN\footnotemark[2] 			& 1.10 & -	& -	\\
	6:5:8:4:8:5 B-B		& $h$-BN\footnotemark[2] 			& 0.79 & -	& -	\\
	6:5:8:4:8:5			& Graphene\footnotemark[2] 		& 0.73 & -	& -	\\			
    \end{tabular}
    \end{ruledtabular}
    \footnotetext{This work}
    \footnotetext{Reference\,\onlinecite{Singh2014}}
  \label{table1}
\end{table}

\begin{figure}[tb!]
\centering
\includegraphics[width=0.85\textwidth]{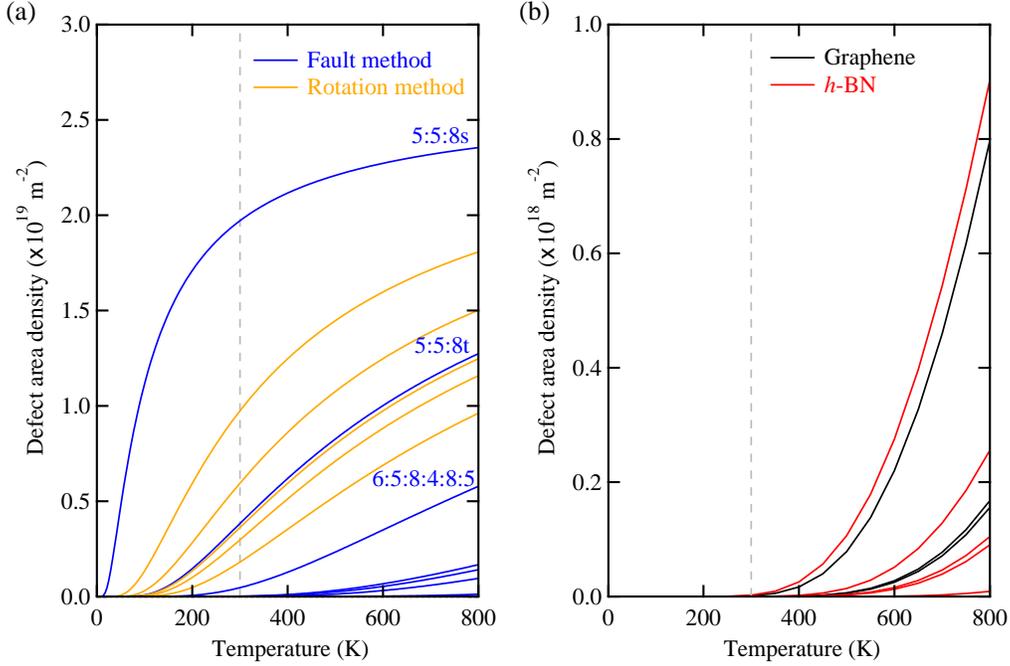} 
\caption{(Color online) Simulated defect area density as a function of temperature: (a) Line defects in phosphorene via the \textit{fault method} (in blue lines, this work) and the \textit{rotation method} (in orange lines, Ref.\,\onlinecite{YLiu2014}), and (b) corresponding fault line defects in graphene\cite{Singh2014} (in black lines) and \textit{h}-BN\cite{Singh2014} (in red lines). Note that the scale is about an order of magnitude smaller in (b)}
\label{figure3}
\end{figure}

To study and determine the relative energetic stability of these 1D line defects in single-layered phosphorene, we define the 1D line defect formation energy per unit length of the line defect, $E^{\rm{LD}}$ using the following equation:
\begin{equation}
	E^{\rm{LD}} = \frac{{E^{\rm{LD}}_{\rm{P}} - N \times E^{\rm{SL}}}}{l} \quad,
\label{eq1}
\end{equation}
where $E^{\rm{LD}}_{\rm{P}}$, $E^{\rm{SL}}$, $N$, and $l$ are the total energy of the single-layered phosphorene with a line defect, the total energy of pristine, defect-free single-layered phosphorene, the number of P atoms in the defect system, and the length of the line defect in the defect system, respectively. Since the basal unit is the length of the line defect, we can think of the calculated $E^{\rm{LD}}$ (in eV/{\AA}) as the additional energy required to form an unit length of line defect for a given pristine single-layered phosphorene.

Now, to collectively address the various line defects in phosphorene, graphene, and $h$-BN under typical growth conditions, we relate the defect area density,\cite{Hu2015} $\rho_{\rm def}$ as a function of temperature, $T$ using the simple Arrhenius equation, 
\begin{equation}
	\rho_{\rm def} = \rho_{\rm pris}\,e^{\tfrac{-E^{\rm D}}{k_{\rm B}T}} \quad.
	\label{eq2}
\end{equation}
Here, the atom area density of the defect-free pristine 2D material is represented by $\rho_{\rm pris}$ and is found to be $2.62\times10^{19}$\,m$^{-2}$, $3.79\times10^{19}$\,m$^{-2}$, and $3.19\times10^{19}$\,m$^{-2}$ for phosphorene, graphene, and $h$-BN, respectively. In this case, the defect formation energy, $E^{\rm D}$ is normalized with respect to number of atoms per line defect length.

As seen in in Tab.\,\ref{table1}, the formation energies of the fault line defects in phosphorene considered in this work are found to be fairly small in magnitude and are also considerably much lower than their counterparts in either $h$-BN or graphene, as reported in Ref.\,\onlinecite{Singh2014}. In line with Ref.\,\onlinecite{YLiu2014}, the energetics of phosphorene grain boundary line defects were calculated to be between 0.05 to 0.14\,eV/{\AA}, lending to the fact that all these 1D defects in phosphorene may have fairly similar chemical stability to defect-free phosphorene than the case for other 2D nanomaterials. Our calculations suggest that many, if not all, of these low energy line defects may very well occur in potential polycrystalline phosphorene-based optoelectronic nanodevices. In comparison to other 2D materials like graphene and $h$-BN (in Fig.\,\ref{figure3}b), their line defects tend to occur only at higher temperatures (above 400\,K) and the corresponding $\rho_{\rm def}$ are almost an order of magntitude less than that of phosphorene line defects.

In fact, uniquely, the 5:5:8s line defect exhibits a remarkably low defect formation energy of 0.01\,eV/{\AA}, i.e. almost {\it close-to-zero} value. In Fig.\,\ref{figure3}a, under typical growth conditions with $T$ between 300 to 800\,K, the 5:5:8s fault line defect structure is predicted to have a very high $\rho_{\rm def}$ as compared to other line defects (including those $rotational$ line defects as reported in Ref.\,\onlinecite{YLiu2014}). Given its slightly bent geometry with P-P dimer pairs along the fault line, we further investigate the effect of lateral strain on the ultra-low energy line defect 5:5:8s by varying the length of the longest edge up to $\pm$\,10\,$\%$. We find that the average P-P bond length, the bent angle, and the total energy of this line defect change negligibly by less than 0.003\,{\AA}, $10^{\circ}$, and 1\,meV per phosphorus atom. This seems to suggest that this ultra-low energy 5:5:8s defect is indeed bent and stable under lateral strain which can become important in actual working nanodevices. 

\section{Electronic structure of one-dimensional line defects}
\begin{figure}[tb!]
\centering
\includegraphics[width=0.75\textwidth]{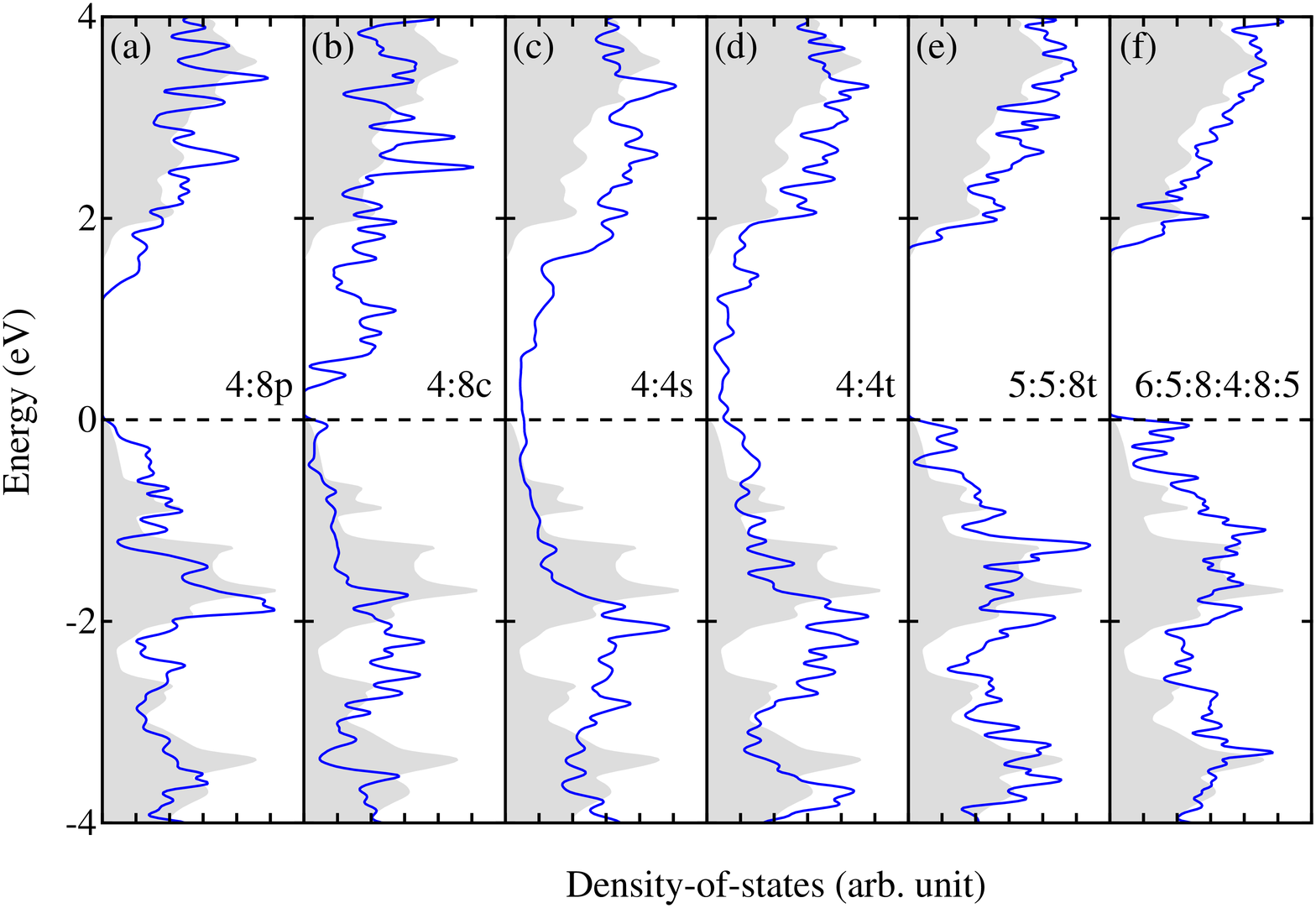} 
\caption{(Color online) Electronic density-of-states (DOS) for (a) 4:8p, (b) 4:8c, (c) 4:4s, (d) 4:4t, (e) 5:5:8t, and (f) 6:5:8:4:8:5 line defect structures. For comparison, the DOS of pristine single-layered phosphorene is indicated by the gray shaded area. Horizontal dotted line at 0\,eV denotes the Fermi energy for metallic structures, and valence band maxium for semiconducting structures.}
\label{figure4}
\end{figure}

Having identified these very low energy 1D line defect structures in single-layered phosphorene, we now turn to their electronic structures. It was recently reported that intrinsic point defects and grain boundaries in single-layered phosphorene are electronically benign/inactive, i.e. preserving the semiconducting nature of phosphorene.\cite{YLiu2014} The authors believe that unlike other heteronuclear 2D nanomaterials (e.g. the metal dichalcogenides), the so-called {\it homoelemental} bonding in phosphorene accounts for this ``absence of chemical disorder", and thus mitigates the formation of deep gap states.\cite{YLiu2014} 

Starting with single-layered defect-free phosphorene, we calculate the total electronic density-of-states (DOS) using the HSE06 hybrid functional, and the $E_{\rm{g}}$ is found to be 1.60\,eV which is in good agreement with other theoretical values.\cite{YLiu2014,VWang2015,Qiao2014} Surprisingly, in contrary to the grain boundary defects studied in Ref.\,\onlinecite{YLiu2014}, we find that the 4:4s, and 4:4t line defect structures exhibit a metallic behavior (Figs.\,\ref{figure4}c and \ref{figure4}d) while 4:8p, 4:8c, 5:5:8s, 5:5:8t, and 6:5:8:4:8:5 yield a semiconducting band gap, $E_{\rm{g}}$ (Figs.\,\ref{figure4}a, \ref{figure4}b, \ref{figure4}e, and \ref{figure4}f). All considered line defect structures doe not possess a net magnetic spin moment.

Taking a closer look at the proposed low energy line defects in this work, we find that those structures containing the square-shaped motifs (hence, 4-fold which deviates from the pristine 3-fold coordination) are metallic (i.e. with no gap). This lack of topological bond preservation (with respect to the 3-fold coordination in pristine phosphorene) explains the observed metallic behavior that was not seen for the grain boundary defects in Ref.\,\onlinecite{YLiu2014}. 

\begin{figure}[tb!]
\centering
\includegraphics[width=0.75\textwidth]{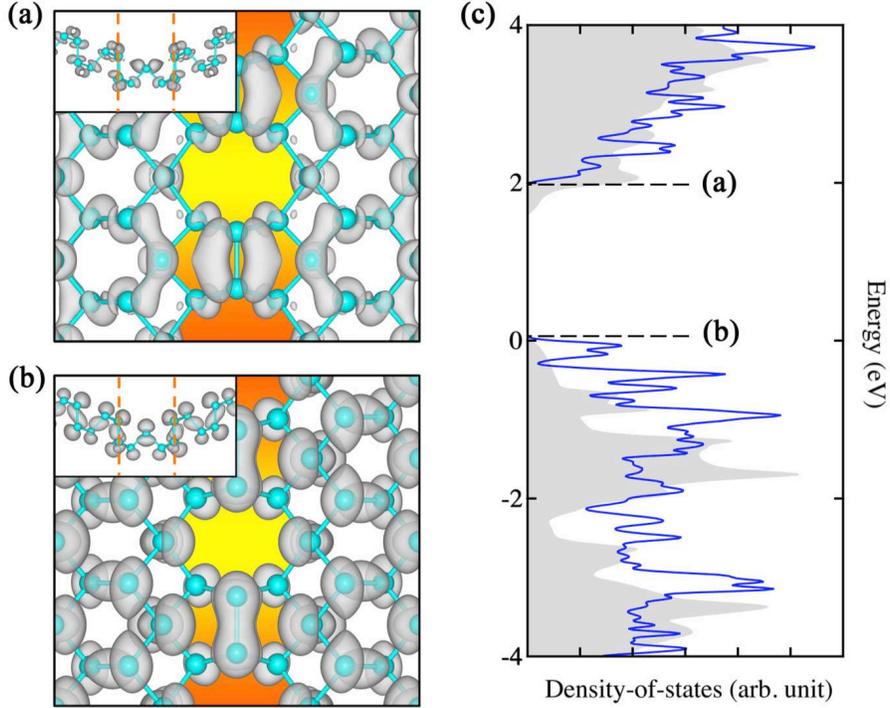} 
\caption{(Color online) Partial electron density and density-of-states of the ultra-low energy 5:5:8s line defect structure. Partial electron density is plotted at the isosurface value of $2\times10^{-4}$\,e/{\AA}$^3$, for both the (a) conduction band and the (b) valence band edges.}
\label{figure5}
\end{figure}

On the other hand, those defect structures preserving the 3-fold P centers (including those with pentagon-shaped motifs) are indeed ``electronically benign/inactive", maintaining a fairly similar band gap energy with pristine phosphorene. Notwithstanding, the remarkably low energy defect 5:5:8s structure is found to have a calculated $E_{\rm{g}}$ of 2.03\,eV -- somewhat unexpectedly wider than that of defect-free phosphorene. We rationalize the increase in the $E_{\rm{g}}$ to the newly formed localized defect states near the valence band edge, as seen in Fig.\,\ref{figure5}c, as a direct consequence of the structural modifications and atomic rearrangements along the defect fault lines.

As can be seen from Tab.\,\ref{table1}, the metallic line defect structures do tend to have a slightly higher line defect formation energy than that of the semiconducting ones. At a first glance, these low energy line defects in single-layered phosphorene may provide a clue into the possibility of modulating the electronic properties by carefully engineering these line defects for new optoelectronic nanodevices.

To further understand the origin of the newly formed defect states near the conduction band minimum and the valence band maximum, we plot the partial electron densities of the ultra-low energy line defect 5:5:8s structure in Figs.\,\ref{figure5}a and \ref{figure5}b, respectively. Comparing to pristine phosphorene where each P atom forms covalent bonds with three neighboring P atoms via the $3p$ electrons, the overall chemical bonding character of this defect structure is follows the similar inter-mixing of the P $3p$ orbitals. 

Specifically, near the conduction band edge (cf. Fig.\,\ref{figure5}a), the partial electron density is found to be largely located between the P atoms located in the same plane (i.e. along the ``{\it xy}" direction) and a $\pi^*$-like orbital character at the P-P dimer along the fault line. However, near the valence band edge (cf. Fig.\,\ref{figure5}b), the partial electron density is largely confined between the P atoms in different planes (i.e. along the ``{\it z}" direction) with a $\sigma$-like bond forming between the P atoms in the P-P dimer parallel to the fault line.

\begin{figure}[tb!]
\centering
\includegraphics[width=0.55\textwidth]{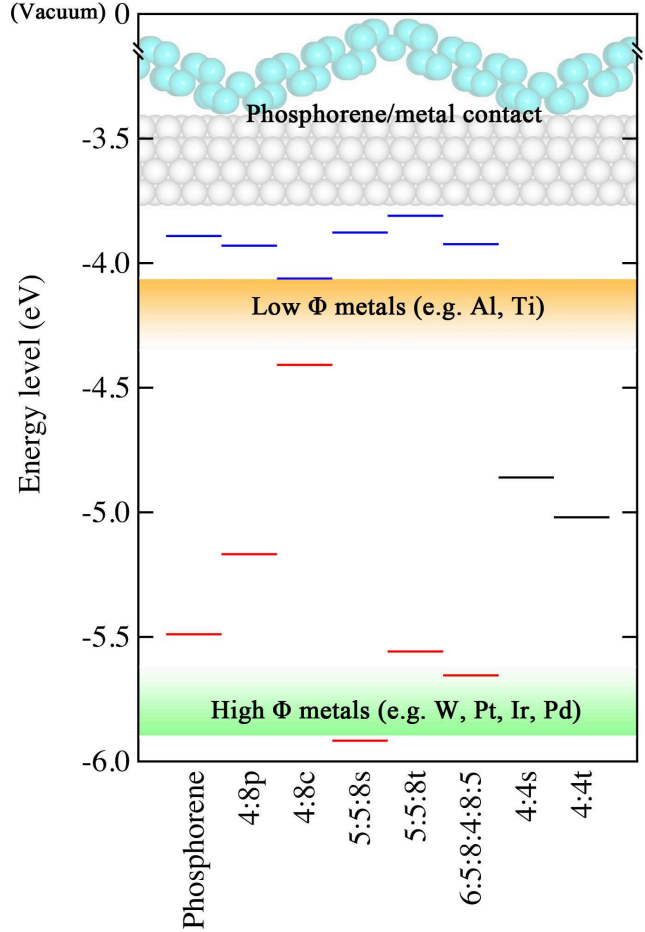}
\caption{(Color online) HSE06-predicted band alignment between phosphorene (with fault line defects) and various metal contacts. The valence band maximum (VBM) and conduction band minimum (CBM) of the semiconducting defects are shown in red and blue horizontal bars, respectively while the Fermi level of the metallic defects are shown in black. The work functions of the various metals are taken from Refs.\,\onlinecite{Holzl1979,Riviere1969,Michaelson1977,Skriver1992}.}
\label{figure6} 
\end{figure}

\section{Phosphorene/metal contacts: Band gap alignment}
It is known that controlling the relative positions of the band edges (with respect to the Fermi level of the metal contact) is crucial to effectively design and tailor the electronic properties in nanoscale FET devices. Recently, it was found that varying the number of phosphorene layers in face-contact with various metal electrodes drastically modulates the position of the valence band maximum and thus modifies the Schottky barrier height.\cite{Cai2014} To compare, we calculate the corresponding work functions of the fault line defect structures and align their band edges with respect to the vacuum level, as shown in Fig.\,\ref{figure6}. Next, we benchmark the relative band edge positions with the work functions of various commonly-used metal contacts (as obtained from Refs.\,\onlinecite{Holzl1979,Riviere1969,Michaelson1977,Skriver1992}).

In particular, the low-energy line defect 5:5:8s is found to shift valence band edge to lower energies by almost 0.5\,eV, as compared to that of pristine phosphorene. On the other hand, the next low-energy line defect 5:5:8t moves the conduction band edge to higher energies, and thus slightly broadens energy band gap. Along these lines, by carefully applying line defect engineering in phosphorene, the generation of selected low-energy line defect structures might act as an effective Ohmic contact with certain metals of high work function values such as tungsten, platinum, irridium, and palladium. This atomic-scale control directly allows one to extend the tunability of $p$-type phosphorene-based FET device performance.

\section{Summary}
To conclude, by using the fault method, various structures of phosphorene with 1D fault line defects were designed and studied using first-principles DFT calculations. We investigated the energetics and electronic structure of these fault line defects, and identified new low-energy fault line defects (i.e. even much lower fault defect formation energy when compared to those found on graphene and $h$-BN). We also showed how these low-energy fault line defects could exhibit a range of electronic structure (from metallic to semiconducting). The lowest fault line defect is termed 5:5:8s which comprises P-P dimer pairs along the fault line, and this new ultra-low energy defect structure (with a formation energy of 0.01\,eV/{\AA}) is bent at a small angle of $136.35^{\circ}$ with a energy band gap of 2.03\,eV. We propose that many, if not all, of these 1D line defects may co-exist during large-scale synthesis of polycrystalline 2D phosphorene, even under growth conditions. Finally, our calculations extends our understanding of the impact of 1D fault line defects in $p$-type phosphorene-based FET nanodevices, going beyond that of the commonly studied point defects.\\

\begin{acknowledgments}
We gratefully acknowledge support from the Basic Science Research Program by the NRF (Grant No. 2014R1A1A1003415). Computational resources have been provided by the KISTI supercomputing center (KSC-2015-C3-009).
\end{acknowledgments}


\begin{thebibliography}{35}%
\makeatletter
\providecommand \@ifxundefined [1]{%
 \@ifx{#1\undefined}
}%
\providecommand \@ifnum [1]{%
 \ifnum #1\expandafter \@firstoftwo
 \else \expandafter \@secondoftwo
 \fi
}%
\providecommand \@ifx [1]{%
 \ifx #1\expandafter \@firstoftwo
 \else \expandafter \@secondoftwo
 \fi
}%
\providecommand \natexlab [1]{#1}%
\providecommand \enquote  [1]{``#1''}%
\providecommand \bibnamefont  [1]{#1}%
\providecommand \bibfnamefont [1]{#1}%
\providecommand \citenamefont [1]{#1}%
\providecommand \href@noop [0]{\@secondoftwo}%
\providecommand \href [0]{\begingroup \@sanitize@url \@href}%
\providecommand \@href[1]{\@@startlink{#1}\@@href}%
\providecommand \@@href[1]{\endgroup#1\@@endlink}%
\providecommand \@sanitize@url [0]{\catcode `\\12\catcode `\$12\catcode
  `\&12\catcode `\#12\catcode `\^12\catcode `\_12\catcode `\%12\relax}%
\providecommand \@@startlink[1]{}%
\providecommand \@@endlink[0]{}%
\providecommand \url  [0]{\begingroup\@sanitize@url \@url }%
\providecommand \@url [1]{\endgroup\@href {#1}{\urlprefix }}%
\providecommand \urlprefix  [0]{URL }%
\providecommand \Eprint [0]{\href }%
\providecommand \doibase [0]{http://dx.doi.org/}%
\providecommand \selectlanguage [0]{\@gobble}%
\providecommand \bibinfo  [0]{\@secondoftwo}%
\providecommand \bibfield  [0]{\@secondoftwo}%
\providecommand \translation [1]{[#1]}%
\providecommand \BibitemOpen [0]{}%
\providecommand \bibitemStop [0]{}%
\providecommand \bibitemNoStop [0]{.\EOS\space}%
\providecommand \EOS [0]{\spacefactor3000\relax}%
\providecommand \BibitemShut  [1]{\csname bibitem#1\endcsname}%
\let\auto@bib@innerbib\@empty
\bibitem [{\citenamefont {Zou}\ and\ \citenamefont
  {Yakobson}(2015{\natexlab{a}})}]{Zou2014}%
  \BibitemOpen
  \bibfield  {author} {\bibinfo {author} {\bibfnamefont {X.}~\bibnamefont
  {Zou}}\ and\ \bibinfo {author} {\bibfnamefont {B.~I.}\ \bibnamefont
  {Yakobson}},\ }\href@noop {} {\bibfield  {journal} {\bibinfo  {journal} {Acc.
  Chem. Res.}\ }\textbf {\bibinfo {volume} {48}},\ \bibinfo {pages} {73}
  (\bibinfo {year} {2015}{\natexlab{a}})}\BibitemShut {NoStop}%
\bibitem [{\citenamefont {Zou}\ and\ \citenamefont
  {Yakobson}(2015{\natexlab{b}})}]{Zou2015}%
  \BibitemOpen
  \bibfield  {author} {\bibinfo {author} {\bibfnamefont {X.}~\bibnamefont
  {Zou}}\ and\ \bibinfo {author} {\bibfnamefont {B.~I.}\ \bibnamefont
  {Yakobson}},\ }\href@noop {} {\bibfield  {journal} {\bibinfo  {journal}
  {Small}\ }\textbf {\bibinfo {volume} {11}},\ \bibinfo {pages} {4503}
  (\bibinfo {year} {2015}{\natexlab{b}})}\BibitemShut {NoStop}%
\bibitem [{\citenamefont {Wang}\ \emph
  {et~al.}(2015{\natexlab{a}})\citenamefont {Wang}, \citenamefont {Li},\ and\
  \citenamefont {Liu}}]{JWang2015}%
  \BibitemOpen
  \bibfield  {author} {\bibinfo {author} {\bibfnamefont {J.}~\bibnamefont
  {Wang}}, \bibinfo {author} {\bibfnamefont {S.~N.}\ \bibnamefont {Li}}, \ and\
  \bibinfo {author} {\bibfnamefont {J.~B.}\ \bibnamefont {Liu}},\ }\href@noop
  {} {\bibfield  {journal} {\bibinfo  {journal} {J. Phys. Chem. A}\ }\textbf
  {\bibinfo {volume} {119}},\ \bibinfo {pages} {3621} (\bibinfo {year}
  {2015}{\natexlab{a}})}\BibitemShut {NoStop}%
\bibitem [{\citenamefont {Carlsson}\ \emph {et~al.}(2011)\citenamefont
  {Carlsson}, \citenamefont {Ghiringhelli},\ and\ \citenamefont
  {Fasolino}}]{Carlsson2011}%
  \BibitemOpen
  \bibfield  {author} {\bibinfo {author} {\bibfnamefont {J.~M.}\ \bibnamefont
  {Carlsson}}, \bibinfo {author} {\bibfnamefont {L.~M.}\ \bibnamefont
  {Ghiringhelli}}, \ and\ \bibinfo {author} {\bibfnamefont {A.}~\bibnamefont
  {Fasolino}},\ }\href@noop {} {\bibfield  {journal} {\bibinfo  {journal}
  {Phys. Rev. B}\ }\textbf {\bibinfo {volume} {84}},\ \bibinfo {pages} {165423}
  (\bibinfo {year} {2011})}\BibitemShut {NoStop}%
\bibitem [{\citenamefont {Zhou}\ \emph {et~al.}(2013)\citenamefont {Zhou},
  \citenamefont {Zou}, \citenamefont {Najmaei}, \citenamefont {Liu},
  \citenamefont {Kong}, \citenamefont {Lou}, \citenamefont {Ajayan},
  \citenamefont {Yakobson},\ and\ \citenamefont {Idrobo}}]{Zhou2013}%
  \BibitemOpen
  \bibfield  {author} {\bibinfo {author} {\bibfnamefont {W.}~\bibnamefont
  {Zhou}}, \bibinfo {author} {\bibfnamefont {X.}~\bibnamefont {Zou}}, \bibinfo
  {author} {\bibfnamefont {S.}~\bibnamefont {Najmaei}}, \bibinfo {author}
  {\bibfnamefont {Z.}~\bibnamefont {Liu}}, \bibinfo {author} {\bibfnamefont
  {J.}~\bibnamefont {Kong}}, \bibinfo {author} {\bibfnamefont {J.}~\bibnamefont
  {Lou}}, \bibinfo {author} {\bibfnamefont {P.~M.}\ \bibnamefont {Ajayan}},
  \bibinfo {author} {\bibfnamefont {B.~I.}\ \bibnamefont {Yakobson}}, \ and\
  \bibinfo {author} {\bibfnamefont {J.-C.}\ \bibnamefont {Idrobo}},\
  }\href@noop {} {\bibfield  {journal} {\bibinfo  {journal} {Nano Lett.}\
  }\textbf {\bibinfo {volume} {13}},\ \bibinfo {pages} {2615} (\bibinfo {year}
  {2013})}\BibitemShut {NoStop}%
\bibitem [{\citenamefont {Kalantar-Zadeh}\ \emph {et~al.}(2012)\citenamefont
  {Kalantar-Zadeh}, \citenamefont {Kis}, \citenamefont {Coleman},\ and\
  \citenamefont {Strano}}]{Wang2012}%
  \BibitemOpen
  \bibfield  {author} {\bibinfo {author} {\bibfnamefont {Q.~H. W.~K.}\
  \bibnamefont {Kalantar-Zadeh}}, \bibinfo {author} {\bibfnamefont
  {A.}~\bibnamefont {Kis}}, \bibinfo {author} {\bibfnamefont {J.~N.}\
  \bibnamefont {Coleman}}, \ and\ \bibinfo {author} {\bibfnamefont {M.~S.}\
  \bibnamefont {Strano}},\ }\href@noop {} {\bibfield  {journal} {\bibinfo
  {journal} {Nat. Nanotechnol.}\ }\textbf {\bibinfo {volume} {7}},\ \bibinfo
  {pages} {699} (\bibinfo {year} {2012})}\BibitemShut {NoStop}%
\bibitem [{\citenamefont {Novoselov}\ \emph {et~al.}(2004)\citenamefont
  {Novoselov}, \citenamefont {Geim}, \citenamefont {Morozov}, \citenamefont
  {Jiang}, \citenamefont {Zhang}, \citenamefont {Dubonos}, \citenamefont
  {Grigorieva},\ and\ \citenamefont {Firsov}}]{Novoselov2004}%
  \BibitemOpen
  \bibfield  {author} {\bibinfo {author} {\bibfnamefont {K.~S.}\ \bibnamefont
  {Novoselov}}, \bibinfo {author} {\bibfnamefont {A.~K.}\ \bibnamefont {Geim}},
  \bibinfo {author} {\bibfnamefont {S.~V.}\ \bibnamefont {Morozov}}, \bibinfo
  {author} {\bibfnamefont {D.}~\bibnamefont {Jiang}}, \bibinfo {author}
  {\bibfnamefont {Y.}~\bibnamefont {Zhang}}, \bibinfo {author} {\bibfnamefont
  {S.~V.}\ \bibnamefont {Dubonos}}, \bibinfo {author} {\bibfnamefont {I.~V.}\
  \bibnamefont {Grigorieva}}, \ and\ \bibinfo {author} {\bibfnamefont {A.~A.}\
  \bibnamefont {Firsov}},\ }\href@noop {} {\bibfield  {journal} {\bibinfo
  {journal} {Science}\ }\textbf {\bibinfo {volume} {306}},\ \bibinfo {pages}
  {666} (\bibinfo {year} {2004})}\BibitemShut {NoStop}%
\bibitem [{\citenamefont {Novoselov}\ \emph {et~al.}(2005)\citenamefont
  {Novoselov}, \citenamefont {Jiang}, \citenamefont {Schedin}, \citenamefont
  {Booth}, \citenamefont {Khotkevich}, \citenamefont {Morozov},\ and\
  \citenamefont {Geim}}]{Novoselov2005}%
  \BibitemOpen
  \bibfield  {author} {\bibinfo {author} {\bibfnamefont {K.~S.}\ \bibnamefont
  {Novoselov}}, \bibinfo {author} {\bibfnamefont {D.}~\bibnamefont {Jiang}},
  \bibinfo {author} {\bibfnamefont {F.}~\bibnamefont {Schedin}}, \bibinfo
  {author} {\bibfnamefont {T.~J.}\ \bibnamefont {Booth}}, \bibinfo {author}
  {\bibfnamefont {V.~V.}\ \bibnamefont {Khotkevich}}, \bibinfo {author}
  {\bibfnamefont {S.~V.}\ \bibnamefont {Morozov}}, \ and\ \bibinfo {author}
  {\bibfnamefont {A.~K.}\ \bibnamefont {Geim}},\ }\href@noop {} {\bibfield
  {journal} {\bibinfo  {journal} {Proc. Natl. Acad. Sci. U.S.A.}\ }\textbf
  {\bibinfo {volume} {102}},\ \bibinfo {pages} {10451} (\bibinfo {year}
  {2005})}\BibitemShut {NoStop}%
\bibitem [{\citenamefont {Jariwala}\ \emph {et~al.}(2014)\citenamefont
  {Jariwala}, \citenamefont {Sangwan}, \citenamefont {Lauhon}, \citenamefont
  {Marks},\ and\ \citenamefont {Hersam}}]{Jariwala2014}%
  \BibitemOpen
  \bibfield  {author} {\bibinfo {author} {\bibfnamefont {D.}~\bibnamefont
  {Jariwala}}, \bibinfo {author} {\bibfnamefont {K.~K.}\ \bibnamefont
  {Sangwan}}, \bibinfo {author} {\bibfnamefont {L.~J.}\ \bibnamefont {Lauhon}},
  \bibinfo {author} {\bibfnamefont {T.~J.}\ \bibnamefont {Marks}}, \ and\
  \bibinfo {author} {\bibfnamefont {M.~C.}\ \bibnamefont {Hersam}},\
  }\href@noop {} {\bibfield  {journal} {\bibinfo  {journal} {ACS Nano}\
  }\textbf {\bibinfo {volume} {8}},\ \bibinfo {pages} {1102} (\bibinfo {year}
  {2014})}\BibitemShut {NoStop}%
\bibitem [{\citenamefont {Schwierz}\ \emph {et~al.}(2015)\citenamefont
  {Schwierz}, \citenamefont {Pezoldt},\ and\ \citenamefont
  {Granzner}}]{Schwierz2015}%
  \BibitemOpen
  \bibfield  {author} {\bibinfo {author} {\bibfnamefont {F.}~\bibnamefont
  {Schwierz}}, \bibinfo {author} {\bibfnamefont {J.}~\bibnamefont {Pezoldt}}, \
  and\ \bibinfo {author} {\bibfnamefont {R.}~\bibnamefont {Granzner}},\
  }\href@noop {} {\bibfield  {journal} {\bibinfo  {journal} {Nanoscale}\
  }\textbf {\bibinfo {volume} {7}},\ \bibinfo {pages} {8261} (\bibinfo {year}
  {2015})}\BibitemShut {NoStop}%
\bibitem [{\citenamefont {Li}\ \emph {et~al.}(2014)\citenamefont {Li},
  \citenamefont {Yu}, \citenamefont {Ye}, \citenamefont {Ge}, \citenamefont
  {Ou}, \citenamefont {Wu}, \citenamefont {Feng}, \citenamefont {Chen},\ and\
  \citenamefont {Zhang}}]{Li2014}%
  \BibitemOpen
  \bibfield  {author} {\bibinfo {author} {\bibfnamefont {L.}~\bibnamefont
  {Li}}, \bibinfo {author} {\bibfnamefont {Y.}~\bibnamefont {Yu}}, \bibinfo
  {author} {\bibfnamefont {G.~J.}\ \bibnamefont {Ye}}, \bibinfo {author}
  {\bibfnamefont {Q.}~\bibnamefont {Ge}}, \bibinfo {author} {\bibfnamefont
  {X.}~\bibnamefont {Ou}}, \bibinfo {author} {\bibfnamefont {H.}~\bibnamefont
  {Wu}}, \bibinfo {author} {\bibfnamefont {D.}~\bibnamefont {Feng}}, \bibinfo
  {author} {\bibfnamefont {X.~H.}\ \bibnamefont {Chen}}, \ and\ \bibinfo
  {author} {\bibfnamefont {Y.}~\bibnamefont {Zhang}},\ }\href@noop {}
  {\bibfield  {journal} {\bibinfo  {journal} {Nat. Nanotechnol.}\ }\textbf
  {\bibinfo {volume} {9}},\ \bibinfo {pages} {372} (\bibinfo {year}
  {2014})}\BibitemShut {NoStop}%
\bibitem [{\citenamefont {Liu}\ \emph {et~al.}(2014{\natexlab{a}})\citenamefont
  {Liu}, \citenamefont {Neal}, \citenamefont {Zhu}, \citenamefont {Luo},
  \citenamefont {Xu}, \citenamefont {Tom\'{a}nek},\ and\ \citenamefont
  {Ye}}]{HLiu2014}%
  \BibitemOpen
  \bibfield  {author} {\bibinfo {author} {\bibfnamefont {H.}~\bibnamefont
  {Liu}}, \bibinfo {author} {\bibfnamefont {A.~T.}\ \bibnamefont {Neal}},
  \bibinfo {author} {\bibfnamefont {Z.}~\bibnamefont {Zhu}}, \bibinfo {author}
  {\bibfnamefont {Z.}~\bibnamefont {Luo}}, \bibinfo {author} {\bibfnamefont
  {X.}~\bibnamefont {Xu}}, \bibinfo {author} {\bibfnamefont {D.}~\bibnamefont
  {Tom\'{a}nek}}, \ and\ \bibinfo {author} {\bibfnamefont {P.~D.}\ \bibnamefont
  {Ye}},\ }\href@noop {} {\bibfield  {journal} {\bibinfo  {journal} {ACS Nano}\
  }\textbf {\bibinfo {volume} {8}},\ \bibinfo {pages} {4033} (\bibinfo {year}
  {2014}{\natexlab{a}})}\BibitemShut {NoStop}%
\bibitem [{\citenamefont {Liu}\ \emph {et~al.}(2014{\natexlab{b}})\citenamefont
  {Liu}, \citenamefont {Xu}, \citenamefont {Zhang}, \citenamefont {Penev},\
  and\ \citenamefont {Yakobson}}]{YLiu2014}%
  \BibitemOpen
  \bibfield  {author} {\bibinfo {author} {\bibfnamefont {Y.}~\bibnamefont
  {Liu}}, \bibinfo {author} {\bibfnamefont {F.}~\bibnamefont {Xu}}, \bibinfo
  {author} {\bibfnamefont {Z.}~\bibnamefont {Zhang}}, \bibinfo {author}
  {\bibfnamefont {E.~S.}\ \bibnamefont {Penev}}, \ and\ \bibinfo {author}
  {\bibfnamefont {B.~I.}\ \bibnamefont {Yakobson}},\ }\href@noop {} {\bibfield
  {journal} {\bibinfo  {journal} {Nano Lett.}\ }\textbf {\bibinfo {volume}
  {14}},\ \bibinfo {pages} {6782} (\bibinfo {year}
  {2014}{\natexlab{b}})}\BibitemShut {NoStop}%
\bibitem [{\citenamefont {Brent}\ \emph {et~al.}(2014)\citenamefont {Brent},
  \citenamefont {Savijani}, \citenamefont {Lewis}, \citenamefont {Haigh},
  \citenamefont {Lewis},\ and\ \citenamefont {O'Brien}}]{Brent2014}%
  \BibitemOpen
  \bibfield  {author} {\bibinfo {author} {\bibfnamefont {J.~R.}\ \bibnamefont
  {Brent}}, \bibinfo {author} {\bibfnamefont {N.}~\bibnamefont {Savijani}},
  \bibinfo {author} {\bibfnamefont {E.~A.}\ \bibnamefont {Lewis}}, \bibinfo
  {author} {\bibfnamefont {S.~J.}\ \bibnamefont {Haigh}}, \bibinfo {author}
  {\bibfnamefont {D.~J.}\ \bibnamefont {Lewis}}, \ and\ \bibinfo {author}
  {\bibfnamefont {P.}~\bibnamefont {O'Brien}},\ }\href@noop {} {\bibfield
  {journal} {\bibinfo  {journal} {Chem. Commun.}\ }\textbf {\bibinfo {volume}
  {50}},\ \bibinfo {pages} {13338} (\bibinfo {year} {2014})}\BibitemShut
  {NoStop}%
\bibitem [{\citenamefont {Buscema}\ \emph {et~al.}(2014)\citenamefont
  {Buscema}, \citenamefont {Groenendijk}, \citenamefont {Blanter},
  \citenamefont {Steele}, \citenamefont {van~der Zant},\ and\ \citenamefont
  {Castellanos-Gomez}}]{Buscema2014}%
  \BibitemOpen
  \bibfield  {author} {\bibinfo {author} {\bibfnamefont {M.}~\bibnamefont
  {Buscema}}, \bibinfo {author} {\bibfnamefont {D.~J.}\ \bibnamefont
  {Groenendijk}}, \bibinfo {author} {\bibfnamefont {S.~I.}\ \bibnamefont
  {Blanter}}, \bibinfo {author} {\bibfnamefont {G.~A.}\ \bibnamefont {Steele}},
  \bibinfo {author} {\bibfnamefont {H.~S.~J.}\ \bibnamefont {van~der Zant}}, \
  and\ \bibinfo {author} {\bibfnamefont {A.}~\bibnamefont
  {Castellanos-Gomez}},\ }\href@noop {} {\bibfield  {journal} {\bibinfo
  {journal} {Nano Lett.}\ }\textbf {\bibinfo {volume} {14}},\ \bibinfo {pages}
  {3347} (\bibinfo {year} {2014})}\BibitemShut {NoStop}%
\bibitem [{\citenamefont {Koenig}\ \emph {et~al.}(2014)\citenamefont {Koenig},
  \citenamefont {Doganov}, \citenamefont {Schmidt}, \citenamefont {Neto},\ and\
  \citenamefont {\"{O}zyilmaz}}]{Koenig2014}%
  \BibitemOpen
  \bibfield  {author} {\bibinfo {author} {\bibfnamefont {S.~P.}\ \bibnamefont
  {Koenig}}, \bibinfo {author} {\bibfnamefont {R.~A.}\ \bibnamefont {Doganov}},
  \bibinfo {author} {\bibfnamefont {H.}~\bibnamefont {Schmidt}}, \bibinfo
  {author} {\bibfnamefont {A.~H.~C.}\ \bibnamefont {Neto}}, \ and\ \bibinfo
  {author} {\bibfnamefont {B.}~\bibnamefont {\"{O}zyilmaz}},\ }\href@noop {}
  {\bibfield  {journal} {\bibinfo  {journal} {Appl. Phys. Lett.}\ }\textbf
  {\bibinfo {volume} {104}},\ \bibinfo {pages} {103106} (\bibinfo {year}
  {2014})}\BibitemShut {NoStop}%
\bibitem [{\citenamefont {Cai}\ \emph {et~al.}(2014)\citenamefont {Cai},
  \citenamefont {Zhang},\ and\ \citenamefont {Zhang}}]{Cai2014}%
  \BibitemOpen
  \bibfield  {author} {\bibinfo {author} {\bibfnamefont {Y.}~\bibnamefont
  {Cai}}, \bibinfo {author} {\bibfnamefont {G.}~\bibnamefont {Zhang}}, \ and\
  \bibinfo {author} {\bibfnamefont {Y.-W.}\ \bibnamefont {Zhang}},\ }\href@noop
  {} {\bibfield  {journal} {\bibinfo  {journal} {Sci. Rep.}\ }\textbf {\bibinfo
  {volume} {4}},\ \bibinfo {pages} {6677} (\bibinfo {year} {2014})}\BibitemShut
  {NoStop}%
\bibitem [{\citenamefont {Kou}\ \emph {et~al.}(2015)\citenamefont {Kou},
  \citenamefont {Chen},\ and\ \citenamefont {Smith}}]{Kou2015}%
  \BibitemOpen
  \bibfield  {author} {\bibinfo {author} {\bibfnamefont {L.}~\bibnamefont
  {Kou}}, \bibinfo {author} {\bibfnamefont {C.}~\bibnamefont {Chen}}, \ and\
  \bibinfo {author} {\bibfnamefont {S.~C.}\ \bibnamefont {Smith}},\ }\href@noop
  {} {\bibfield  {journal} {\bibinfo  {journal} {J. Phys. Chem. Lett.}\
  }\textbf {\bibinfo {volume} {6}},\ \bibinfo {pages} {2794} (\bibinfo {year}
  {2015})}\BibitemShut {NoStop}%
\bibitem [{\citenamefont {Wei}\ and\ \citenamefont {Peng}(2014)}]{Wei2014}%
  \BibitemOpen
  \bibfield  {author} {\bibinfo {author} {\bibfnamefont {Q.}~\bibnamefont
  {Wei}}\ and\ \bibinfo {author} {\bibfnamefont {X.}~\bibnamefont {Peng}},\
  }\href@noop {} {\bibfield  {journal} {\bibinfo  {journal} {Appl. Phys.
  Lett.}\ }\textbf {\bibinfo {volume} {104}},\ \bibinfo {pages} {251915}
  (\bibinfo {year} {2014})}\BibitemShut {NoStop}%
\bibitem [{\citenamefont {Singh}\ and\ \citenamefont
  {Waghmare}(2014)}]{Singh2014}%
  \BibitemOpen
  \bibfield  {author} {\bibinfo {author} {\bibfnamefont {A.}~\bibnamefont
  {Singh}}\ and\ \bibinfo {author} {\bibfnamefont {U.~V.}\ \bibnamefont
  {Waghmare}},\ }\href@noop {} {\bibfield  {journal} {\bibinfo  {journal}
  {Phys. Chem. Chem. Phys.}\ }\textbf {\bibinfo {volume} {16}},\ \bibinfo
  {pages} {21664} (\bibinfo {year} {2014})}\BibitemShut {NoStop}%
\bibitem [{\citenamefont {Liu}\ \emph {et~al.}(2012)\citenamefont {Liu},
  \citenamefont {Zou},\ and\ \citenamefont {Yakobson}}]{Liu2012}%
  \BibitemOpen
  \bibfield  {author} {\bibinfo {author} {\bibfnamefont {Y.}~\bibnamefont
  {Liu}}, \bibinfo {author} {\bibfnamefont {X.}~\bibnamefont {Zou}}, \ and\
  \bibinfo {author} {\bibfnamefont {B.~I.}\ \bibnamefont {Yakobson}},\
  }\href@noop {} {\bibfield  {journal} {\bibinfo  {journal} {ACS Nano}\
  }\textbf {\bibinfo {volume} {6}},\ \bibinfo {pages} {7053} (\bibinfo {year}
  {2012})}\BibitemShut {NoStop}%
\bibitem [{\citenamefont {Kahaly}\ \emph {et~al.}(2008)\citenamefont {Kahaly},
  \citenamefont {Singh},\ and\ \citenamefont {Waghmare}}]{Kahaly2008}%
  \BibitemOpen
  \bibfield  {author} {\bibinfo {author} {\bibfnamefont {M.~U.}\ \bibnamefont
  {Kahaly}}, \bibinfo {author} {\bibfnamefont {S.~P.}\ \bibnamefont {Singh}}, \
  and\ \bibinfo {author} {\bibfnamefont {U.~V.}\ \bibnamefont {Waghmare}},\
  }\href@noop {} {\bibfield  {journal} {\bibinfo  {journal} {Small}\ }\textbf
  {\bibinfo {volume} {4}},\ \bibinfo {pages} {2209} (\bibinfo {year}
  {2008})}\BibitemShut {NoStop}%
\bibitem [{\citenamefont {Kresse}\ and\ \citenamefont
  {Hafner}(1993)}]{Kresse1993}%
  \BibitemOpen
  \bibfield  {author} {\bibinfo {author} {\bibfnamefont {G.}~\bibnamefont
  {Kresse}}\ and\ \bibinfo {author} {\bibfnamefont {J.}~\bibnamefont
  {Hafner}},\ }\href@noop {} {\bibfield  {journal} {\bibinfo  {journal} {Phys.
  Rev. B}\ }\textbf {\bibinfo {volume} {47}},\ \bibinfo {pages} {558} (\bibinfo
  {year} {1993})}\BibitemShut {NoStop}%
\bibitem [{\citenamefont {Kresse}\ and\ \citenamefont
  {Hafner}(1994)}]{Kresse1994}%
  \BibitemOpen
  \bibfield  {author} {\bibinfo {author} {\bibfnamefont {G.}~\bibnamefont
  {Kresse}}\ and\ \bibinfo {author} {\bibfnamefont {J.}~\bibnamefont
  {Hafner}},\ }\href@noop {} {\bibfield  {journal} {\bibinfo  {journal} {Phys.
  Rev. B}\ }\textbf {\bibinfo {volume} {49}},\ \bibinfo {pages} {14251}
  (\bibinfo {year} {1994})}\BibitemShut {NoStop}%
\bibitem [{\citenamefont {Kresse}\ and\ \citenamefont
  {Furthm\"{u}ller}(1996)}]{Kresse1996}%
  \BibitemOpen
  \bibfield  {author} {\bibinfo {author} {\bibfnamefont {G.}~\bibnamefont
  {Kresse}}\ and\ \bibinfo {author} {\bibfnamefont {J.}~\bibnamefont
  {Furthm\"{u}ller}},\ }\href@noop {} {\bibfield  {journal} {\bibinfo
  {journal} {Phys. Rev. B}\ }\textbf {\bibinfo {volume} {54}},\ \bibinfo
  {pages} {11169} (\bibinfo {year} {1996})}\BibitemShut {NoStop}%
\bibitem [{\citenamefont {Kresse}\ and\ \citenamefont
  {Joubert}(1999)}]{Kresse1999}%
  \BibitemOpen
  \bibfield  {author} {\bibinfo {author} {\bibfnamefont {G.}~\bibnamefont
  {Kresse}}\ and\ \bibinfo {author} {\bibfnamefont {D.}~\bibnamefont
  {Joubert}},\ }\href@noop {} {\bibfield  {journal} {\bibinfo  {journal} {Phys.
  Rev. B}\ }\textbf {\bibinfo {volume} {59}},\ \bibinfo {pages} {1758}
  (\bibinfo {year} {1999})}\BibitemShut {NoStop}%
\bibitem [{\citenamefont {Perdew}\ \emph {et~al.}(1996)\citenamefont {Perdew},
  \citenamefont {Burke},\ and\ \citenamefont {Ernzerhof}}]{Perdew1996}%
  \BibitemOpen
  \bibfield  {author} {\bibinfo {author} {\bibfnamefont {J.~P.}\ \bibnamefont
  {Perdew}}, \bibinfo {author} {\bibfnamefont {K.}~\bibnamefont {Burke}}, \
  and\ \bibinfo {author} {\bibfnamefont {M.}~\bibnamefont {Ernzerhof}},\
  }\href@noop {} {\bibfield  {journal} {\bibinfo  {journal} {Phys. Rev. Lett.}\
  }\textbf {\bibinfo {volume} {77}},\ \bibinfo {pages} {3865} (\bibinfo {year}
  {1996})}\BibitemShut {NoStop}%
\bibitem [{\citenamefont {Yazyev}\ and\ \citenamefont
  {Louie}(2010)}]{Yazyev2010}%
  \BibitemOpen
  \bibfield  {author} {\bibinfo {author} {\bibfnamefont {O.~V.}\ \bibnamefont
  {Yazyev}}\ and\ \bibinfo {author} {\bibfnamefont {S.~G.}\ \bibnamefont
  {Louie}},\ }\href@noop {} {\bibfield  {journal} {\bibinfo  {journal} {Phys.
  Rev. B}\ }\textbf {\bibinfo {volume} {81}},\ \bibinfo {pages} {195420}
  (\bibinfo {year} {2010})}\BibitemShut {NoStop}%
\bibitem [{\citenamefont {Hu}\ and\ \citenamefont {Yang}(2015)}]{Hu2015}%
  \BibitemOpen
  \bibfield  {author} {\bibinfo {author} {\bibfnamefont {W.}~\bibnamefont
  {Hu}}\ and\ \bibinfo {author} {\bibfnamefont {J.}~\bibnamefont {Yang}},\
  }\href@noop {} {\bibfield  {journal} {\bibinfo  {journal} {J. Phys. Chem. C}\
  }\textbf {\bibinfo {volume} {119}},\ \bibinfo {pages} {20474} (\bibinfo
  {year} {2015})}\BibitemShut {NoStop}%
\bibitem [{\citenamefont {Wang}\ \emph
  {et~al.}(2015{\natexlab{b}})\citenamefont {Wang}, \citenamefont {Kawazoe},\
  and\ \citenamefont {Geng}}]{VWang2015}%
  \BibitemOpen
  \bibfield  {author} {\bibinfo {author} {\bibfnamefont {V.}~\bibnamefont
  {Wang}}, \bibinfo {author} {\bibfnamefont {Y.}~\bibnamefont {Kawazoe}}, \
  and\ \bibinfo {author} {\bibfnamefont {W.~T.}\ \bibnamefont {Geng}},\
  }\href@noop {} {\bibfield  {journal} {\bibinfo  {journal} {Phys. Rev. B}\
  }\textbf {\bibinfo {volume} {91}},\ \bibinfo {pages} {045433} (\bibinfo
  {year} {2015}{\natexlab{b}})}\BibitemShut {NoStop}%
\bibitem [{\citenamefont {Qiao}\ \emph {et~al.}(2014)\citenamefont {Qiao},
  \citenamefont {Kong}, \citenamefont {Hu}, \citenamefont {Yang},\ and\
  \citenamefont {Ji}}]{Qiao2014}%
  \BibitemOpen
  \bibfield  {author} {\bibinfo {author} {\bibfnamefont {J.}~\bibnamefont
  {Qiao}}, \bibinfo {author} {\bibfnamefont {X.}~\bibnamefont {Kong}}, \bibinfo
  {author} {\bibfnamefont {Z.-X.}\ \bibnamefont {Hu}}, \bibinfo {author}
  {\bibfnamefont {F.}~\bibnamefont {Yang}}, \ and\ \bibinfo {author}
  {\bibfnamefont {W.}~\bibnamefont {Ji}},\ }\href@noop {} {\bibfield  {journal}
  {\bibinfo  {journal} {Nat. Commun.}\ }\textbf {\bibinfo {volume} {5}},\
  \bibinfo {pages} {4475} (\bibinfo {year} {2014})}\BibitemShut {NoStop}%
\bibitem [{\citenamefont {H\"{o}lzl}\ \emph {et~al.}(1979)\citenamefont
  {H\"{o}lzl}, \citenamefont {Schulte},\ and\ \citenamefont
  {Wagner}}]{Holzl1979}%
  \BibitemOpen
  \bibfield  {author} {\bibinfo {author} {\bibfnamefont {J.}~\bibnamefont
  {H\"{o}lzl}}, \bibinfo {author} {\bibfnamefont {F.~K.}\ \bibnamefont
  {Schulte}}, \ and\ \bibinfo {author} {\bibfnamefont {H.}~\bibnamefont
  {Wagner}},\ }\href@noop {} {\emph {\bibinfo {title} {Solid Surface
  Physics}}}\ (\bibinfo  {publisher} {Springer-Verlag, Berlin},\ \bibinfo
  {year} {1979})\ Chap.\ \bibinfo {chapter} {Work Functions of
  Metals}\BibitemShut {NoStop}%
\bibitem [{\citenamefont {Riviere}(1969)}]{Riviere1969}%
  \BibitemOpen
  \bibfield  {author} {\bibinfo {author} {\bibfnamefont {J.~C.}\ \bibnamefont
  {Riviere}},\ }\href@noop {} {\emph {\bibinfo {title} {Solid State Surface
  Science}}},\ Vol.~\bibinfo {volume} {1}\ (\bibinfo  {publisher} {Dekker, New
  York},\ \bibinfo {year} {1969})\ Chap.\ \bibinfo {chapter} {Work Function:
  Measurements and Results}\BibitemShut {NoStop}%
\bibitem [{\citenamefont {Michaelson}(1977)}]{Michaelson1977}%
  \BibitemOpen
  \bibfield  {author} {\bibinfo {author} {\bibfnamefont {H.~B.}\ \bibnamefont
  {Michaelson}},\ }\href@noop {} {\bibfield  {journal} {\bibinfo  {journal} {J.
  Appl. Phys.}\ }\textbf {\bibinfo {volume} {48}},\ \bibinfo {pages} {4729}
  (\bibinfo {year} {1977})}\BibitemShut {NoStop}%
\bibitem [{\citenamefont {Skriver}\ and\ \citenamefont
  {Rosengaard}(1992)}]{Skriver1992}%
  \BibitemOpen
  \bibfield  {author} {\bibinfo {author} {\bibfnamefont {H.~L.}\ \bibnamefont
  {Skriver}}\ and\ \bibinfo {author} {\bibfnamefont {N.~M.}\ \bibnamefont
  {Rosengaard}},\ }\href@noop {} {\bibfield  {journal} {\bibinfo  {journal}
  {Phys. Rev. B}\ }\textbf {\bibinfo {volume} {46}},\ \bibinfo {pages} {7157}
  (\bibinfo {year} {1992})}\BibitemShut {NoStop}%
\end{thebibliography}
%

\end{document}